\documentclass[fleqn,usenatbib]{mnras}

\usepackage{hyperref}
\usepackage{epsfig}
\usepackage{myaasmacros}
\usepackage{amsmath}
\usepackage{amsfonts}
\usepackage{subfig}
\usepackage{comment}
\usepackage{color}

\title[Nature versus nurture] 
{Nature versus nurture: what regulates star formation in satellite galaxies?}  
  
\author[De Lucia et al.]{Gabriella De Lucia$^{1}$\thanks{E-mail:
    gabriella.delucia@inaf.it}, Michaela Hirschmann$^{2}$,
  Fabio Fontanot$^{1}$\\
$^{1}$INAF - Astronomical Observatory of Trieste, via G.B. Tiepolo 11,
  I-34143 Trieste, Italy\\
$^{2}$UPMC-CNRS, UMR7095, Institut d'~Astrophysique de Paris, F-75014
Paris, France}

\begin{document}

\date{Accepted ???. Received ??? in original form ???}

\pagerange{\pageref{firstpage}--\pageref{lastpage}} \pubyear{2018}

\maketitle

\label{firstpage}

\begin{abstract}
  We use our state-of-the-art Galaxy Evolution and Assembly (GAEA)
  semi-analytic model to study how and on which time-scales star formation is
  suppressed in satellite galaxies. Our fiducial stellar feedback model,
  implementing strong stellar driven outflows, reproduces relatively well the
  variations of passive fractions as a function of galaxy stellar mass and halo
  mass measured in the local Universe, as well as the `quenching' time-scales
  inferred from the data. We show that the same level of agreement can be
  obtained by using an alternative stellar feedback scheme featuring lower
  ejection rates at high redshift, and modifying the treatment for hot gas
  stripping. This scheme over-predicts the number densities of low to
    intermediate mass galaxies. In addition, a good agreement with the observed
    passive fractions can be obtained only by assuming that cooling can
    continue on satellites, at the rate predicted considering halo properties
    at infall, even after their parent dark matter substructure is stripped
    below the resolution of the simulation. For our fiducial model, the better
    agreement with the observed passive fractions can be ascribed to: (i) a
  larger cold gas fraction of satellites at the time of accretion, and (ii) a
  lower rate of gas reheating by supernovae explosions and stellar winds with
  respect to previous versions of our model. Our results suggest that the
  abundance of passive galaxies with stellar mass larger than $\sim
  10^{10}\,{\rm M}_{\sun}$ is primarily determined by the self-regulation
  between star formation and stellar feedback, with environmental processes
  playing a more marginal role.
\end{abstract}


\section{Introduction}\label{intro}
It has long been known that the physical properties of galaxies exhibit strong
correlations with the environment they live in. Galaxies in over-dense regions
have typically more evolved (elliptical) morphologies, lower gas masses, redder
colours, and lower star formation rates than galaxies of similar mass living in
regions of the Universe with mean density
\citep[e.g.][]{Oemler_1974,Dressler_1980,Giovanelli_Haynes_1983}. In more
recent years, studies on the subject have received much impetus from the
completion of large spectroscopic surveys both in the local Universe and at
higher redshift \citep[][just to name a
  few]{Kauffmann_etal_2004,Balogh_etal_2004,Cooper_etal_2012,Fossati_etal_2017,Cucciati_etal_2017}.

At z=0, a detailed characterization of the role of environment on the physical
properties of galaxies has been obtained using group/cluster catalogues based
on data from the Sloan Digital Sky Survey (SDSS). These studies have shown that
the fraction of red/passive galaxies increases with increasing galaxy stellar
mass, parent halo mass, and local density, and that the colour/star formation
activity distribution is bimodal across all range of halo masses/environments
probed
\citep[e.g.][]{Weinmann_etal_2006,Baldry_etal_2006,Kimm_etal_2009,Peng_etal_2010,DeLucia_etal_2012,Wetzel_etal_2012,Hirschmann_etal_2014}. These
observations are often interpreted in a `satellite versus centrals framework':
galaxies located at the centre of the dominant halo are defined as `central'
galaxies, while the others are classified as `satellites'. While this scheme is
intuitive and straightforward from the theoretical point of view, its
application to observational data is generally more difficult and typically
requires mock simulated catalogues for calibration\footnote{For a discussion
  about the applicability of the framework, and about its general relevance in
  studies of galaxy evolution, we refer to
  \citet*{DeLucia_Muzzin_Weinmann_2014}}. Adopting this interpretative scheme,
observational measurements in the local Universe have been used to constrain
the efficiency and time-scales of physical processes affecting the star
formation activity of satellite galaxies.  \citet{DeLucia_etal_2012} combined
galaxy merger trees extracted from semi-analytic models with measurements of
the red/passive fractions of satellites in groups/clusters, and estimated
quenching time-scales varying between 5 and 7 Gyr.  \citet{Wetzel_etal_2013}
constrained satellite star formation histories combining infall histories
extracted from high-resolution N-body simulations with measurements of the
specific star formation rate (i.e. the ratio between the star formation rate
and the galaxy stellar mass, ${\rm SFR/M_{star}}$) distribution. They put
forward a so called `delayed-then-rapid quenching' scenario for satellite
galaxies: the specific SFRs are unaffected for 2-4 Gyr after infall, and then
decay on a very short time-scale (e-folding time $< 0.8$~Gyr).
\citet{Hirschmann_etal_2014} revisited the analysis by
\citet{DeLucia_etal_2012} extending it to the dependence of the quiescent
fraction on local galaxy density, and demonstrated that the inferred long
quenching time-scales were difficult to achieve in the framework of the back
then state-of-the-art semi-analytic models. The studies just mentioned all
focused on galaxies with stellar mass larger than $\sim 10^{9.5}\,{\rm
  M}_{\sun}$. At lower galaxy stellar masses ($<10^8\,{\rm M}_{\sun}$) the
quenching efficiency appears to increase dramatically, possibly as a
consequence of the increasing importance of ram-pressure stripping
\citep{Fillingham_etal_2015,Fillingham_etal_2016}.

From a theoretical standpoint, central galaxies occupy a `special' location
where cold gas accretion can take place either rapidly along filamentary
structures at high redshift, or more gradually, cooling from a quasi-static hot
atmosphere, at low redshift and in more massive haloes. When a galaxy is
accreted onto a larger system (i.e. it becomes a satellite), its hot gas
reservoir is typically assumed to be instantaneously and completely removed, so
that no new replenishment of the cold gas is possible through gas cooling. This
process is usually referred to as `strangulation' or `starvation', and
simulation studies have argued that a non negligible amount of hot gas can
remain in place even several Gyrs after accretion
\citep{McCarthy_etal_2008}. In addition, a number of different physical
processes can effectively reduce the cold gas content of satellite
galaxies. For example, satellite galaxies travelling at high velocities through
a dense intra-cluster medium suffer a strong ram-pressure stripping that can
sweep cold gas out of the stellar discs \citep{Gunn_and_Gott_1972}. Repeated
high-speed encounters with other satellite galaxies
\citep{Farouki_and_Shapiro_1981,Moore_etal_1996} and, more generally,
galaxy-galaxy interactions \citep[e.g.][]{Mihos_2004} can lead to a strong
internal dynamical response, triggering the conversion of part of the available
cold gas into stars. In modern theoretical models of galaxy formation, these
physical processes are typically combined with efficient stellar feedback
schemes, which translates in a rather rapid depletion of the cold gas reservoir
associated with satellite galaxies, and an excess of red/passive galaxies with
respect to the observations \citep[e.g.][]{Weinmann_etal_2006b,Wang_etal_2007}.

When this problem was first noted, about one decade ago, the attention of the
community focused on the `simplified' treatment of satellite galaxies and, in
particular, on the assumption that the hot gas reservoir is instantaneously
stripped at the accretion time. In later studies, based on semi-analytic
models, a more gradual stripping of the hot gas has been assumed. Albeit
improved, however, the agreement with observational measurements has been far
from satisfactory
\citep[e.g.][]{Kang_vdBosch_2008,Font_etal_2008,Weinmann_etal_2010,Guo_etal_2011,Hirschmann_etal_2014}. The
problem has not been limited to semi-analytic models
\citep{Weinmann_etal_2012}, and it still appears in state-of-the-art
hydrodynamical simulations. For example, in the Hydrangea simulations
\citep{Bahe_etal_2017}, that adopt the galaxy formation model developed for the
EAGLE project, the fraction of passive satellites {\it decreases} with
increasing stellar mass, which is opposite to what is observed. In addition,
the fraction of passive satellites predicted by these simulations is
significantly larger than observed for intermediate to low mass satellites (see
Fig.~6 in \citealt{Bahe_etal_2017}). A comparison with predictions from the
Illustris simulation was published in \citet{Sales_etal_2015}, who analysed the
projected radial distribution of satellites split by colour, and found a good
agreement with observational measurements based on SDSS. The agreement was
ascribed to the large (relative to earlier work) gas contents of satellite
galaxies at the time of infall. In fact, \citet{Sales_etal_2015} show (see
their Fig.~3) that the gas fractions of infalling satellites in Illustris are
significantly larger than those predicted by the semi-analytic model by
\citet{Guo_etal_2011}, and in quite good agreement with local observations of
HI in galaxies. While it would be useful to see a more detailed comparison
between these simulations (as well as their next generation -
\citealt{Pillepich_etal_2018}) and the measured variations of passive fractions
as a function of halo and galaxy mass, the nature of the simulations makes it
difficult (and certainly computationally very expensive) to understand exactly
{\it why} (i.e. through which physical processes) a better agreement with
observational results can be achieved. The flexibility of the semi-analytic
approach is an advantage in this case, as it allows us to disentangle
efficiently the relative importance of different processes using large
simulated volumes.

A few recent renditions of semi-analytic models exhibit a relatively good
agreement with observational data.  Contrary to what was originally expected,
the success of these models is not driven by an improved satellite treatment:
\citet{Hirschmann_etal_2016} have shown that the predicted variations of
passive fractions as a function of galaxy stellar mass are sensitive to the
adopted stellar feedback scheme. In addition, they have been able to identify
schemes that reproduce well the observational measurements, albeit adopting an
instantaneous stripping of the hot gas reservoir at
infall. \citet{Henriques_etal_2017} have used a different approach: they have
re-tuned the parameters of their galaxy formation model to explicitly fit the
measured fractions of passive galaxies. This process required important
modifications for both their star formation and stellar feedback schemes, in
addition to modifying the treatment of hot gas stripping. 

In this paper, we will use our state-of-the-art GAlaxy Evolution and Assembly
({\sc gaea}) semi-analytic model to analyse in detail the time-scales over
which star formation in satellite galaxies is suppressed, and their dependence
on different physical treatments. The layout of the paper is as follows: in
Sections~\ref{theory}, we will introduce the model runs used in this study. In
Section~\ref{passfrac}, we will compare model predictions with the
distributions of specific SFRs and variations of passive fractions as a
function of galaxy stellar mass and halo mass, as measured in the local
Universe. In Section~\ref{qtime}, we will quantify the time-scales over which
the star formation in present-day satellite galaxies is suppressed in our
models. In Section~\ref{discussion} we discuss our results, and compare model
predictions with estimates of passive fractions and quenching time-scales at
earlier cosmic times. Finally, in Section~\ref{concl}, we give our conclusions.

\section{The {\sc Gaea} galaxy formation model}\label{theory}

The {\sc gaea} semi-analytic model builds on the model published in
\citet{DeLucia_and_Blaizot_2007}, but many prescriptions have been updated
significantly over the past years. In particular, the model features a
sophisticated chemical enrichment scheme that accounts for the
non-instantaneous recycling of gas, metals and energy
\citep{DeLucia_etal_2014}, and a modified stellar feedback scheme that is
partly based on results from numerical simulations
\citep{Hirschmann_etal_2016}.

Specifically, we assume that the reheating rate of cold disk gas, due to
supernovae explosions and winds from massive stars, can be parametrized using
the fitting formulae provided by \citet{Muratov_etal_2015}. These are based on
the FIRE (Feedback In Realistic Environments) set of hydrodynamical simulations
\citep{Hopkins_etal_2014}. We also assume that the same parametrizations can be
adopted to model the rate of energy injection in the inter-stellar medium, and
model the gas ejection rate outside galactic haloes following energy
conservation arguments as in \citet{Guo_etal_2011}. Finally, we assume that
ejected gas can be re-accreted onto the hot gas component associated with
central galaxies on a time-scale that depends on halo mass, as in
\citet{Henriques_etal_2013}. For a detailed description of the modelling
adopted for stellar feedback ({\sc hdlf16-fire} hereafter), we refer to
\citet{Hirschmann_etal_2016}. In previous work, we have shown that, in our {\sc
  GAEA} framework, this feedback scheme allows us to reproduce nicely the
fraction of passive galaxies measured in the local Universe, and its variation
as a function of galaxy stellar mass, both for central and satellite
galaxies. In addition, we have also shown that the {\sc hdlf16-fire}
feedback scheme reproduces the observed evolution of the galaxy stellar mass
function and the cosmic star formation rate up to $z\sim 10$, as well as a
number of other important observational constraints
\citep{Hirschmann_etal_2016,Fontanot_etal_2017,Zoldan_etal_2017,Xie_etal_2017,Zoldan_etal_2018}.

In the following sections, we will compare predictions from our fiducial model
with those from a run adopting the `energy-driven' feedback scheme used
in \citet{DeLucia_etal_2004} and \citet{DeLucia_etal_2014}. This model
(hereafter {\sc hdlf16-ed}) assumes that: (i) the reheating rate is
proportional to the inverse of the square virial velocity; (ii) all gas
reheated in central galaxies is ejected, while gas reheated in satellites is
added to the hot gas reservoir associated with centrals; and (iii) ejected gas
is re-incorporated on a halo dynamical time-scale (see first row of
Table~1 in \citealt{Hirschmann_etal_2016}). The {\sc hdlf16-ed} feedback has
been adopted in previous versions of our model, and is meant to illustrate how
much model predictions can vary depending on the assumed stellar feedback
scheme.

Both the {\sc hdlf16-fire} and the {\sc hdlf16-ed} runs assume that the hot gas
halo associated with infalling galaxies is instantaneously stripped at the time
these are accreted onto larger systems. As discussed above, this assumption has
important consequences on the evolution of satellite galaxies. In order to
quantify how much the star formation history of satellites can be affected by a
different treatment for hot gas stripping, we also consider two alternative
runs. These ({\sc ed-nostr1} and {\sc ed-nostr2}) adopt the {\sc hdlf16-ed}
feedback scheme, but relax the instantaneous stripping
assumption. Specifically, satellite galaxies are able to retain their hot gas
halo, which then replenishes the cold disk gas via gas cooling.  We also assume
that the cooling rate on satellite galaxies can be reduced (and eventually
balanced) by radio mode feedback from Active Galactic Nuclei (AGN). For these
two physical processes, we adopt the prescriptions detailed in
\citet{DeLucia_etal_2010} and \citet{Croton_etal_2006}, respectively. We evolve
satellite galaxies using the properties of their parent substructures at the
time of infall (i.e. at the last time these were central haloes). When subhalos
are stripped below the resolution limit of the simulation, our model assumes
that the hosted galaxies survive as `orphans', and merge with the corresponding
central galaxies after a residual merging time based on the Chandrasekar
approximation (see \citealt{DeLucia_and_Blaizot_2007} and
\citealt{DeLucia_etal_2010} for details). In the {\sc ed-nostr1} run, we assume
that the residual hot gas associated with orphan galaxies is stripped and added
to the hot gas reservoir of the parent halo. In an alternative run (hereafter
{\sc ed-nostr2}), we assume that cooling can take place also on orphan
galaxies, until the hot gas reservoir is exhausted, or the galaxy merges. This
is clearly unrealistic, as a significant stripping of the parent subhalo is
bound to lead to some stripping of the baryonic component of the hosted galaxy
(unless subhalo stripping is exclusively numerical). Therefore, this run
provides a lower limit to the fraction of passive satellites that can be
obtained by only modifying the treatment of hot gas stripping, in the
  framework of our {\sc hdlf16-ed} model. In all runs considered, the gas
ejected from satellite galaxies is associated with the ejected reservoir
associated with the corresponding central galaxy. Finally, it is worth
stressing that, with the exception of the {\sc hdlf16-fire} run, all model
variations considered (including the {\sc hdlf16-ed} run) do not reproduce the
measured evolution of the galaxy stellar mass function.

All model predictions shown in the following are based on merger trees
extracted from the Millennium Simulation \citep{Springel_etal_2005}. This is a
pure dark matter $N$-body simulation of a box of $500\,{\rm Mpc}{\rm h}^{-1}$
on a side, assuming cosmological parameters consistent with WMAP1
($\Omega_\Lambda=0.75$, $\Omega_m=0.25$, $\Omega_b=0.045$, $n=1$,
$\sigma_8=0.9$, and $H_0=73 \, {\rm km\,s^{-1}\,Mpc^{-1}}$). Although these
cosmological parameters are now out-to-date (more recent measurements favour
in particular a lower value for $\sigma_8$), previous work has shown that model
results are qualitatively unaffected by relatively small variations of their
values, once the physical parameters of the model are re-tuned to reproduce a
given set of observational results in the local Universe
\citep{Wang_etal_2008,Guo_etal_2013}. All figures presented below have
  been constructed analysing about 20 per cent of the volume of the simulation,
but we have verified that results do not vary significantly when considering a
larger volume.

\section{Passive galaxy fractions in the local Universe}\label{passfrac}

\begin{figure*}
  \centering
  \epsfig{file=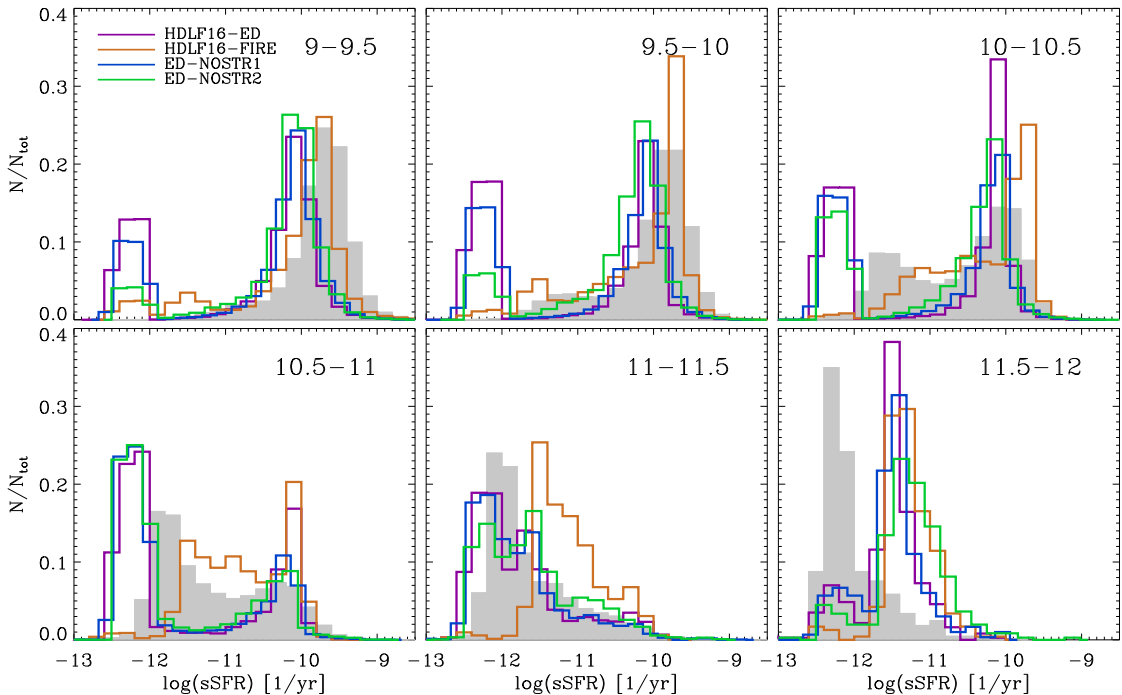, width=0.85\textwidth}
  \caption{Distributions of specific SFRs for different stellar mass bins
    (different panels) at $z=0$, as predicted by the different
    feedback/satellite models considered in this study ({\sc hdlf16-ed}: lila;
    {\sc hdlf16-fire}: orange; {\sc ed-nostr1}: green; {\sc ed-nostr2}:
    red). Model predictions are compared with observational data from SDSS,
    shown by the gray histograms.}
          {\label{sSFRdistr}}
\end{figure*}

Fig.~\ref{sSFRdistr} shows the distributions of specific SFRs for different
galaxy stellar mass bins (different panels, as indicated in the legend). Lines
of different colours correspond to the different runs introduced above, while
gray histograms show observational estimates from \citet{Hirschmann_etal_2014}.
These are based on data from the SDSS DR8, cross-correlated with the JHU-MPA DR7
catalogue\footnote{http://www.mpa-garching.mpg.de/SDSS/DR7/} and the updated
DR7 version of the group catalogue by \citet{Yang_etal_2007}. The specific SFRs
shown are based on measurements published in \citet{Brinchmann_etal_2004},
with modifications regarding the treatment of dust attenuation and aperture
corrections as described in \citet{Salim_etal_2007}. Only model galaxies
brighter than $-18$ in the r-band have been considered, as in the observational
sample. Model galaxies with zero SFR\footnote{The distributions shown here for
  the runs {\sc hdlf16-fire} and {\sc hdlf16-ed} differ from those shown in
  Fig.~8 of \citet{Hirschmann_etal_2016} because galaxies with zero SFR were
  excluded in that work.} are included by assuming specific SFR values
following a Gaussian distribution centred at $12.5\,{\rm yr^{-1}}$, with a
dispersion of $0.25\,{\rm yr^{-1}}$.

None of the models considered here reproduces well the observed distributions
For the lowest mass bin considered, the {\sc hdlf16-fire} run over-predicts
slightly the number of passive galaxies, but extends to specific SFR values
larger than those predicted by the other three runs, and is in better agreement
with the peak corresponding to `active' galaxies in the observed
distribution. For the same stellar mass bin, the {\sc hdlf16-ed} run and its
variants over-predict significantly the fraction of passive galaxies, and
active galaxies peak at a value that is offset low with respect to the
observational measurements. At intermediate galaxy stellar masses, the {\sc
  hdlf16-ed} run and its variants over-predict the numbers of passive galaxies,
but exhibit a clearly bimodal distribution as in the observational data. The
{\sc hdlf16-fire} run performs better in terms of passive fractions, but the
distribution is less bimodal and the passive peak is moved to higher values of
specific SFR with respect to the observations. For the most massive bin
considered, all model distributions are offset towards larger values of
specific SFR with respect to the observed distribution, i.e. model galaxies
tend to be more active than observed. In \citet{Hirschmann_etal_2016}, we
ascribed this failure to the adopted scheme for AGN feedback. We have not been
able, however, to improve the agreement with observational data via a simple
modification of the efficiency of this form of feedback, which suggests the
problem might have a more complex origin and solution.

\begin{figure*}
  \centering
  \epsfig{file=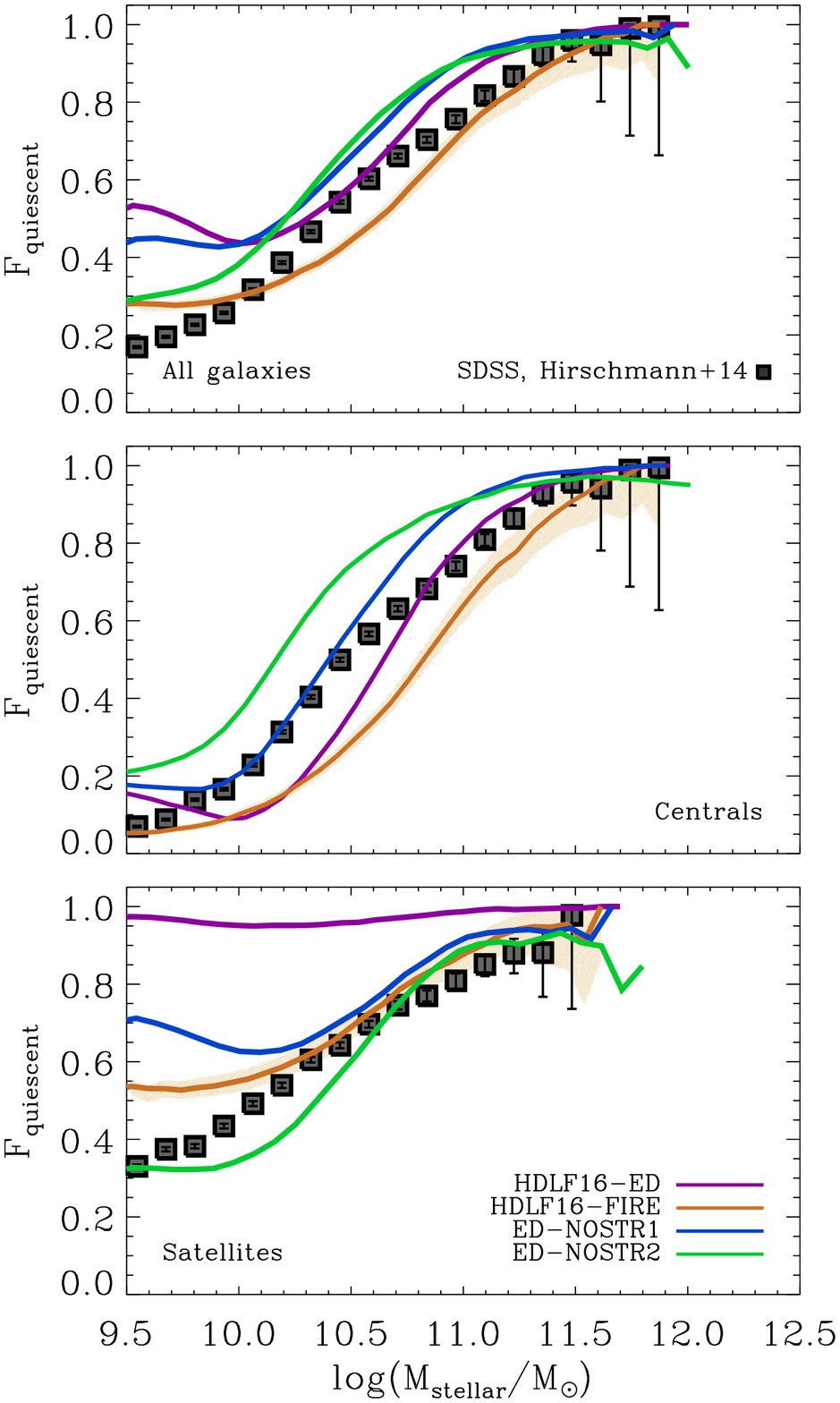,
    width=0.5\textwidth}\hspace{-0.5cm}
   \epsfig{file=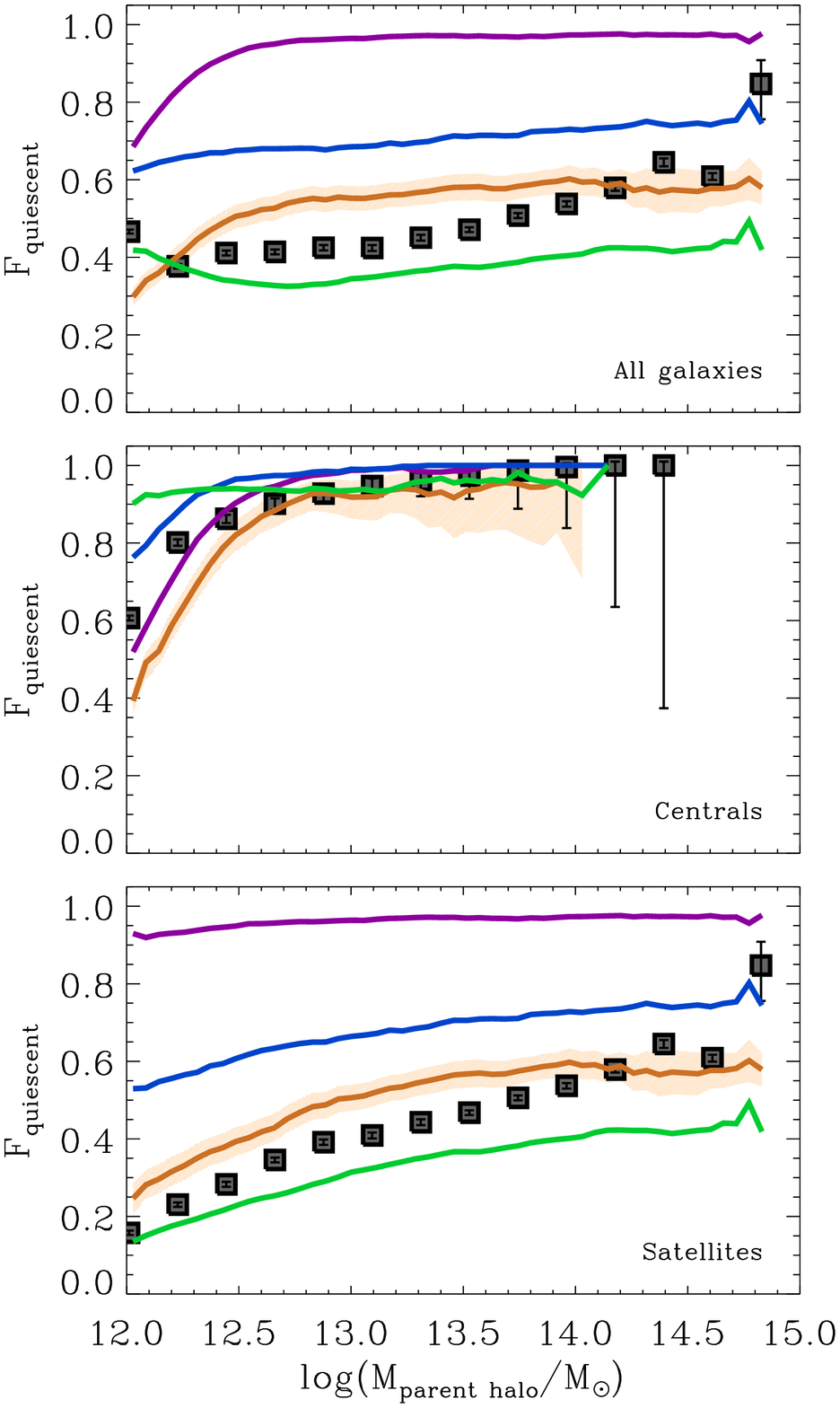,
     width=0.5\textwidth}\hspace{-0.5cm}
  \caption{Present-day quiescent fraction of all galaxies (top row), centrals
    (middle row), and satellites (bottom row) as a function galaxy stellar
    mass (left column) and halo mass (right column). Filled symbols with error
    bars correspond to observational measurements based on the SDSS 
    \citep{Hirschmann_etal_2014}. Lines of different colours correspond to the
    different runs used in this study.}  {\label{Quiescfrac}}
\end{figure*}

In the following analysis, following \citet{Hirschmann_etal_2016}, we select as
passive galaxies all those with specific SFR smaller than $0.3\times t_{\rm
  hubble}^{-1}\approx 10^{-11}\,{\rm yr^{-1}}$ \citep[see
  also][]{Franx_etal_2008}. The model distributions shown in
Fig.~\ref{sSFRdistr} are not in agreement with observational estimates, but are
such that the adopted threshold assures that the predicted passive fractions
are not very sensitive to this mismatch.

Fig.~\ref{Quiescfrac} shows the fraction of passive galaxies as a function of
galaxy stellar mass (left column) and parent halo mass (right column), as
predicted by all model runs considered (lines of different colours), and as
estimated from data (symbols with error bars). Different rows show results for
all galaxies (top), centrals only (middle), and satellites (bottom). The top
panels show that the {\sc hdlf16-fire} model reproduces approximately well the
measured variations of passive fractions both as a function of galaxy stellar
mass and as a function of halo mass. At intermediate galaxy masses and halo
masses, model predictions are slightly below and above the observed values,
respectively. For the lowest galaxy stellar masses considered, our fiducial
model tends to over-predict passive fractions. This could be probably
alleviated by removing the assumption of instantaneous hot gas strippint and by
including an explicit treatment for gas/stellar stripping. All other runs
considered over-predict the fractions of passive galaxies for galaxy masses
below $\sim 10^{10}\,{\rm M}_{\sun}$. As expected, the largest excess is found
for the {\sc hdlf16-ed} run, that also significantly over-predicts the fraction
of passive galaxies measured as a function of halo mass, for the entire mass
range considered. Assuming cooling can continue on satellite galaxies that
retain their parent dark matter substructure ({\sc ed-nostr1}) improves the
agreement with observational measurements, but does not resolve the excess of
low-mass passive galaxies. Only by making the unrealistic assumption that gas
cooling can still continue on orphan galaxies until the hot gas is exhausted
(this is the case in the {\sc ed-nostr2} run), are we able to significantly
reduce the fraction of passive galaxies predicted at low masses. In this run,
the predicted passive fraction as a function of halo mass is even lower than
measured in the data, particularly for the most massive haloes.

The middle panels of Fig.~\ref{Quiescfrac} show the fraction of passive central
galaxies, as predicted and estimated from data.  For virtually all galaxy
stellar masses considered, with the exception of the most massive galaxies,
our fiducial {\sc hdlf16-fire} model under-predicts the measured fraction of
passive centrals, while the {\sc ed-nostr2} tends to over-predict it. The
reason for this is that, in this run, less hot gas is able to cool onto
central galaxies because a certain fraction of it remains associated with
infalling satellites and contributes to replenish their cold gas
reservoir. The over-prediction of passive centrals in this run is such that
there is no significant variation of the central passive fraction as a function
of halo mass (middle right panel). The other runs perform
somewhat better, with the {\sc hdlf16-fire} model under-predicting the fraction
of passive central galaxies for haloes less massive than $\sim 10^{12.5}\,{\rm
  M_{\sun}}$. 

The bottom panels of Fig.~\ref{Quiescfrac} show the fraction of passive
satellite galaxies, as predicted from all runs considered and as measured from
data. The figure clearly shows that virtually all satellites in the {\sc
  hdlf16-ed} run are passive, with very little trend as a function of galaxy
stellar mass. The run also over-predicts by large factors the measured fraction
of passive satellites as a function of halo mass (bottom right panel).
Relaxing the approximation of instantaneous hot gas stripping reduces the
overall fraction of passive satellites and introduces a trend as a function of
galaxy mass. For the lowest galaxy masses considered, the observed passive
fraction can only be reproduced, in the framework of the {\sc hdlf16-ed}
feedback scheme, assuming that gas can cool also on orphan galaxies at the rate
predicted considering halo properties at the time of satellite accretion. The
same run, however, under-predicts the fraction of passive satellites for all
haloes more massive than $\sim 10^{13}\,{\rm M_{\sun}}. $The {\sc hdlf16-fire}
model is in good agreement with observational data for galaxies more massive
than $\sim 10^{10.3}\,{\rm M_{\sun}}$, while it still over-predicts the
fraction of passive satellites for lower galaxy masses.  This indicates the
necessity and importance to improve the adopted treatment for satellite
galaxies at low masses. This could likely also improve the agreement with the
measured fractions of passive satellites as a function of halo mass.

Results shown in Fig.~\ref{Quiescfrac} do not depend on the resolution of the
simulation, but for the {\sc ed-nostr1} model (see Appendix A). In this case, a
higher resolution allows gas cooling to take place on satellites for longer
times (down to lower substructure masses), which makes the passive satellite
fractions lower (closer to those predicted by the {\sc ed-nostr2} model) at low
stellar masses.

In summary, in the framework of the {\sc hdlf16-ed} feedback scheme, the
observed variations of passive fractions as a function of galaxy stellar mass
and halo mass can be reproduced, albeit not perfectly, assuming that the hot
gas reservoir is retained by infalling satellites, and that gas cooling can
take place on them until the hot reservoir is completely exhausted. A
comparable level of agreement with data can be obtained also by assuming a
different feedback scheme, albeit still adopting an instantaneous stripping of
the hot gas reservoir (as in the {\sc hdlf16-fire} run). Since the assumptions
adopted in the former approach are unrealistic, and it does not reproduce the
observed evolution of the galaxy stellar mass function, we favour the {\sc
  hdlf16-fire} model as a better solution to the `over-quenching' problem
discussed earlier. The agreement with data is not perfect in neither cases, but
our exercise demonstrates that the observed variations of passive fractions are
not exclusively determined by environmental effects and can, in fact, be
significantly affected by a modified treatment of stellar feedback.

Finally, we note that a more accurate comparison between model predictions and
observational data would require running the same group finder employed for the
observations on our model catalogues. In fact, \citet{Campbell_etal_2015} have
shown that correlated scatter in galaxy colour at fixed central luminosity and
central/satellite misidentification affect the colour-dependent halo occupation
statistics. While this has an impact on the detailed comparison between
observations and model predictions (particularly for central galaxies), it does
not significantly affect our conclusions.

\section{Quenching time-scales}\label{qtime}

In addition to passive fractions and their trends as a function of galaxy
mass and halo mass, it is also interesting to evaluate the time-scales over
which the star formation in satellite model galaxies is suppressed, and compare
model predictions with available observational estimates. In this section, we
use the `quenching time-scales' estimated by \citet{Wetzel_etal_2013} and the
`refined' estimates by \citet{Hirschmann_etal_2014}. As mentioned in
Section~\ref{intro}, these are obtained combining observational estimates of
the variation of passive fractions and of the specific SFR distributions with
either infall histories extracted from N-body simulations or galaxy merger
trees extracted from semi-analytic models. In
\citet{Hirschmann_etal_2014}, we demonstrated that the `refined' estimates are
consistent with those obtained considering the time elapsed between the last
time each model galaxy was central and star forming, and the first time it
became a satellite passive galaxy. In addition, for this calculation, we only
consider present-day passive satellites that are star-forming at the
time of infall (i.e. were not already quenched before accretion). 

\begin{figure}
  \centering
  \epsfig{file=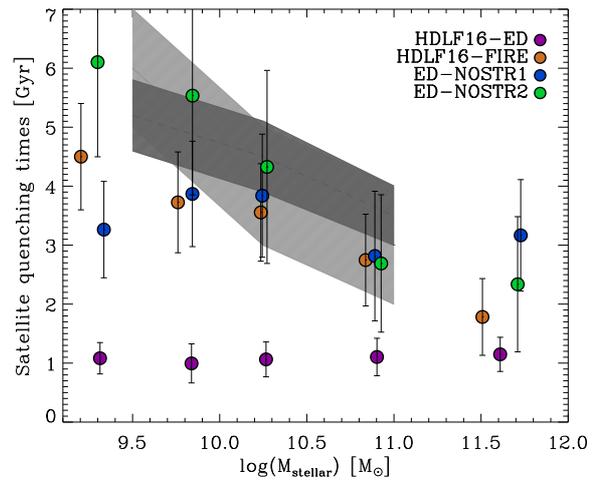,width=0.45\textwidth\hspace{-0.3cm}} 
  \caption{Satellite quenching time-scales versus galaxy stellar mass. The grey
    shaded regions represent observational estimates by
    \citet[][dark grey]{Wetzel_etal_2013} and
    \citet[][light grey]{Hirschmann_etal_2014}. Symbols of different colour
    show model predictions from the four runs considered in this study, as
    indicated in the legend. Only model satellites that are passive at present
    and that were star forming at infall (see text for detail) have been
    considered. For model predictions, we show the mean values of the quenching
    time-scales and their scatter.}  {\label{Quench_times}}
\end{figure}

Fig.~\ref{Quench_times} shows how model predictions from the runs considered
here (circles of different colours) compare to observational estimates. The
quenching time-scales predicted by our {\sc hdlf16-ed} feedback schemes are
very short, of the order of $\sim 1$~Gyr, and do not depend on galaxy stellar
mass. Longer time-scales can be obtained both by relaxing the assumption of
instantaneous hot gas stripping and by modifying the adopted stellar feedback
scheme. The strongest trend as a function of galaxy stellar mass is obtained
for the {\sc ed-nostr2} model, that predicts quenching time-scales as long as
$\sim 6$~Gyrs for galaxies with stellar mass $\sim 9.8\,{\rm M_{\sun}}$. For
galaxies of the same mass, our {\sc hdlf16-fire} model predicts quenching
time-scales that are somewhat lower but still within the uncertainties of the
observational estimates. For the most massive galaxies, where observational
estimates are not available, predictions from the {\sc hdlf16-fire} run are
consistent with those from the {\sc hdlf16-ed} run, and are lower than those
predicted by the other model variants considered.  The figure confirms that the
evolution of satellite galaxies can be influenced by both internal physical
processes (stellar feedback in this case) and by environmental processes. Our
model does not include the effect of ram-pressure stripping of cold gas and/or
tidal stellar stripping. We expect, however, these processes to be more
important in more massive systems and therefore not for the bulk of the
satellite population from the observational samples considered.

\section{Discussion}\label{discussion} 

\begin{figure*}
  \centering
\epsfig{file=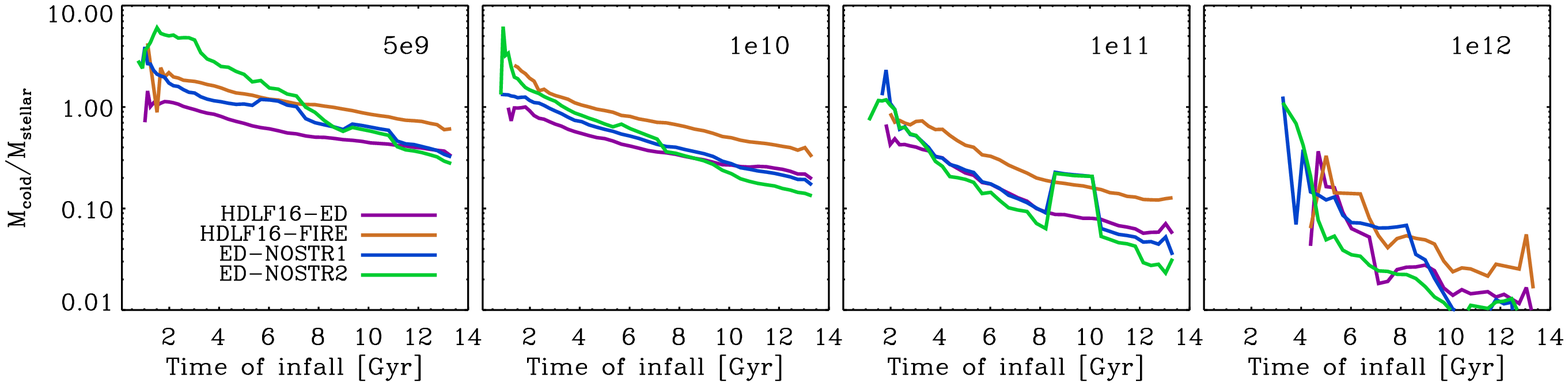,width=0.95\textwidth} 
\epsfig{file=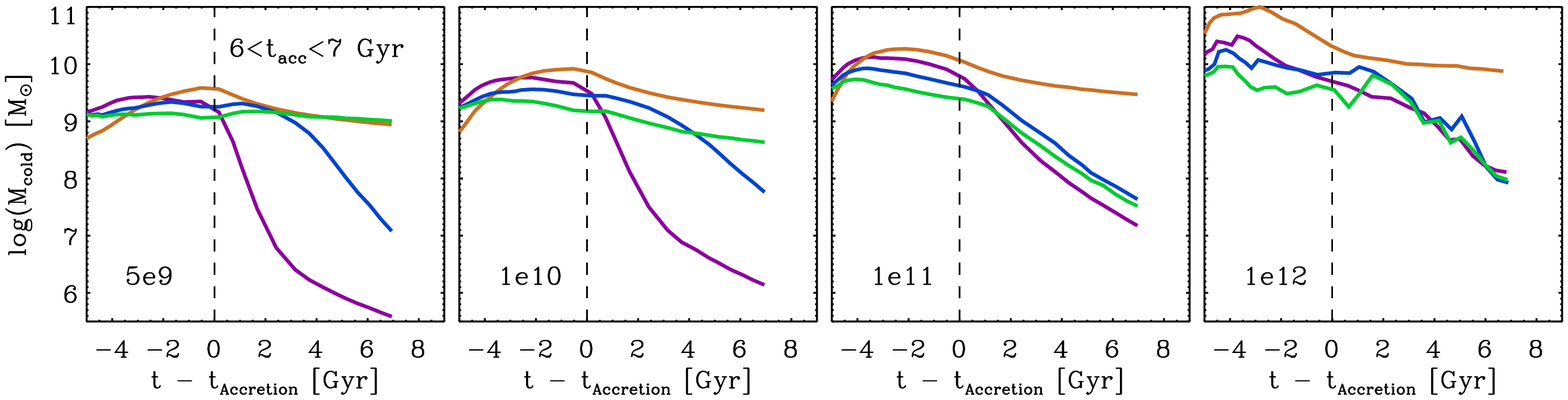,width=0.95\textwidth}
\epsfig{file=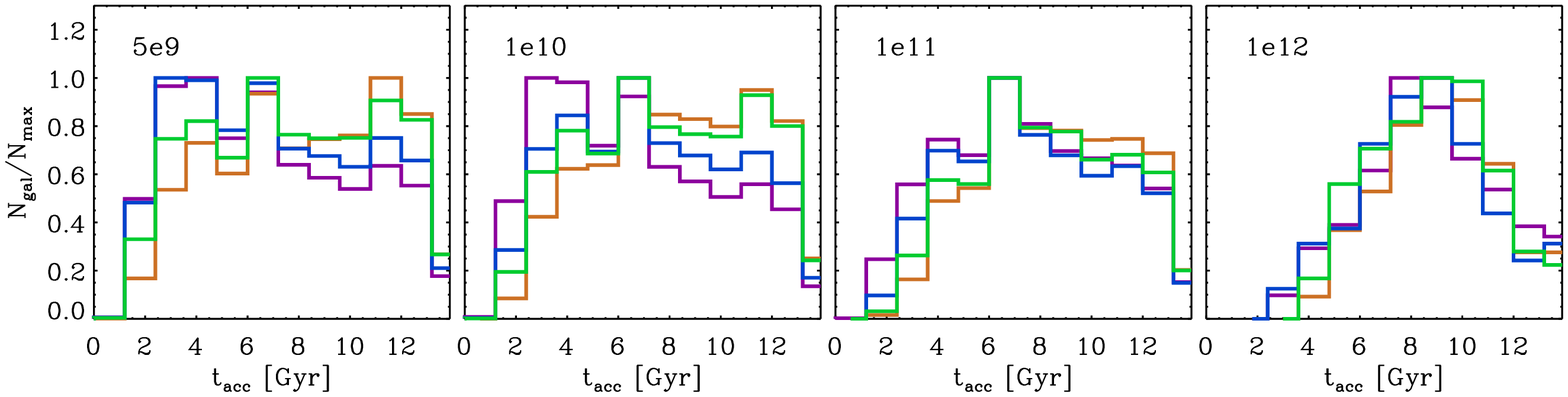,width=0.95\textwidth}
\caption{Top row: cold gas fractions at the time of infall for galaxies that
  are passive satellites at z=0 and that were star forming at the time of
  infall. Middle row: evolution (rescaled to the time of accretion) of the cold
  gas mass associated with present day passive satellites that were accreted
  between 6 and 7 Gyrs ago. Bottom row: distribution of the accretion times of
  the same satellites considered for the top panels. Different columns
  correspond to different galaxy stellar mass bins at $z=0$, as indicated in
  the legend.}  {\label{Evol_Coldgas}}
\end{figure*}

In the previous sections, we have shown that our fiducial {\sc hdlf16-fire}
model is able to reproduce relatively well (albeit not perfectly) the
variations of passive fractions measured in the nearby Universe, as well as the
relatively long quenching time-scales inferred from data. This is somewhat
surprising as our fiducial feedback scheme assumes instantaneous stripping of
the hot gas associated with infalling satellites. Below (in
Section~\ref{sec:why}), we explain in detail how the better agreement with data
is achieved. We have shown that a relatively good agreement with data can be
also found relaxing the assumption of instantaneous hot gas stripping at
infall. In this case, however, it is necessary to assume that gas cooling
continues until exhaustion of the hot gas reservoir, at a rate predicted
considering halo properties at the time of satellite infall, even after the
hosting dark matter substructure has been stripped below the resolution limit
of the simulation. One obvious question is if the same level of agreement is
maintained in both runs at higher redshift and/or if higher redshift data can
be used to discriminate among alternatives models. We will address this
question below (Section~\ref{sec:highz}).

\subsection{The success of our fiducial {\sc hdlf16-fire} feedback scheme}
\label{sec:why}

The top panels of Fig.~\ref{Evol_Coldgas} show the gas fractions at the time of
infall for galaxies that are passive satellites today and were star forming at
the time of infall, in different bins of galaxy stellar mass (different
columns, as indicated by the legend). There is a clear offset between the {\sc
  hdlf16-fire} and {\sc hdlf16-ed} schemes, with the former predicting
systematically higher gas fractions than the latter, for all galaxy masses and
independently of the infall time. The figure also shows that the predicted gas
fractions increase with decreasing galaxy stellar mass, in qualitative
agreement with observational measurements (for a quantitative comparison with
data, see Fig.~5 in \citealt{Hirschmann_etal_2016}). When relaxing the
assumption of instantaneous hot gas stripping, the gas fractions at infall
generally increase with respect to predictions from the {\sc hdlf16-ed} run, as
expected. The increase is modest or negligible for the most massive galaxies,
but becomes more significant at lower galaxy masses: in the {\sc ed-nostr2}
run, satellites of mass $\sim 10^9\,{\rm M_{\sun}}$ accreted earlier than $\sim
8$~Gyr ago have gas fractions that are even {\it larger} than in the {\sc
  hdlf16-fire} run. In addition, the two runs assuming non-instantaneous hot
gas stripping predict a stronger dependence of the cold gas fraction on infall
time with respect to our fiducial and {\sc hdlf16-ed} runs.  Thus, in
principle, measurements of gas fractions as a function of cosmic time could be
used to discriminate between the different runs used here. In practice,
however, the scatter in predicted gas fraction at fixed galaxy stellar mass is
typically rather large, and observational measurements also carry large
uncertainties.

The middle panels of Fig.~\ref{Evol_Coldgas} show the evolution of the cold gas
mass associated with galaxies that are passive satellites today, and that were
accreted between 6 and 7~Gyrs ago. These average trends have been obtained by
linking each model galaxy to its most massive progenitor at each previous
cosmic epoch, and have been rescaled to the time of accretion (i.e. the last
time the galaxy was central). We have verified that the trends remain the same
for earlier/later accretion times, but the relative importance of
external/internal processes varies. The cold gas mass increases at early times
due to cosmological infall, and later decreases due to a combination of star
formation and stellar feedback dominating over gas cooling. For galaxies with
stellar mass larger than $\sim 10^{11}\,{\rm M}_{\sun}$, the decrease of cold
gas, due to star formation, starts before galaxies become satellites. At fixed
galaxy mass today, the cold gas mass at accretion is largest in the {\sc
  hdlf16-fire} run and lowest for the {\sc ed-nostr2} run. After accretion, the
replenishment of new cold gas via cooling is suppressed in the {\sc
  hdlf16-fire} and {\sc hdlf16-ed} runs, and delayed gas recycling from
previous stellar populations does not contribute significant amounts of cold
gas. As a consequence, cold gas always decreases after accretion due to a
combination of star formation and stellar feedback.  The decrease is more rapid
in the {\sc hdlf16-ed} model due to a higher rate of reheating (see Fig. 4 in
\citealt{Hirschmann_etal_2016}). In the {\sc ed-nostr1} and {\sc ed-nostr2}
runs, gas cooling can continue after accretion until the galaxy becomes orphan
in the former run, or until hot gas is exhausted in the latter. For galaxies of
low to intermediate stellar mass, this is enough to prevent a significant
decrease of the cold gas mass for relatively long times. For more massive
galaxies, that have lower gas fractions at the time of accretion (see top
panels), the evolution predicted in these two runs is instead closer to that
predicted by the {\sc hdlf16-ed} model.  Among the four runs considered, the
{\sc hdlf16-fire} exhibits the most rapid increase of cold gas before
accretion, and the slowest depletion at later times.

The bottom panels of Fig.~\ref{Evol_Coldgas} show the distribution of accretion
times for the same satellite galaxies considered in the top panels. Satellites
have been accreted over a range of cosmic epochs, with the most massive
galaxies being accreted on average later than lower mass galaxies
\citep[this is a natural consequence of hierarchical accretion;][]{DeLucia_etal_2012}. Interestingly, we find that galaxies selected in fixed bins of stellar
mass at z=0 correspond to different median accretion times. In particular,
galaxies in the {\sc hdlf16-fire} exhibit the distributions that are most
skewed towards later cosmic times. Therefore, on average, present day passive
satellites in this run have suffered environmental processes for a shorter
time than satellites of the same mass in the other runs. It is worth noting,
however, that gas fractions tend to decrease with cosmic time, and that the
range of accretion times is very large at all stellar masses and in all runs
considered in this study. 

In summary, the better agreement with data of our {\sc hdlf16-fire} run with
respect to predictions from the {\sc hdlf16-ed} run is due to the fact that (i)
galaxies are generally more gas-rich (and therefore more star forming) at
infall, and (ii) the gas re-heating rate assumed in the {\sc hdlf16-fire} run
is lower than that adopted in the {\sc hdlf16-ed} run, allowing satellites
to keep their star-forming reservoir for longer times. 

\subsection{Passive fractions and quenching time-scales at higher redshift}
\label{sec:highz}

The analysis presented in the previous sections is focused on the local
Universe, where detailed measurements of the variations of passive fractions as
a function of galaxy stellar mass and halo mass (or alternative definitions of
the environment) are available. At higher redshifts, statistical studies are
more difficult, due to the more limited availability of spectroscopic redshifts
and overall less accurate measurements of the star formation activity. The
situation is rapidly improving, and will be revolutionized in the next future
with the advent of new space missions like the James Webb Space Telescope and
Euclid and highly multiplexed spectroscopic instruments from the ground (e.g.,
MOONS at VLT).

\begin{figure}
  \centering
\epsfig{file=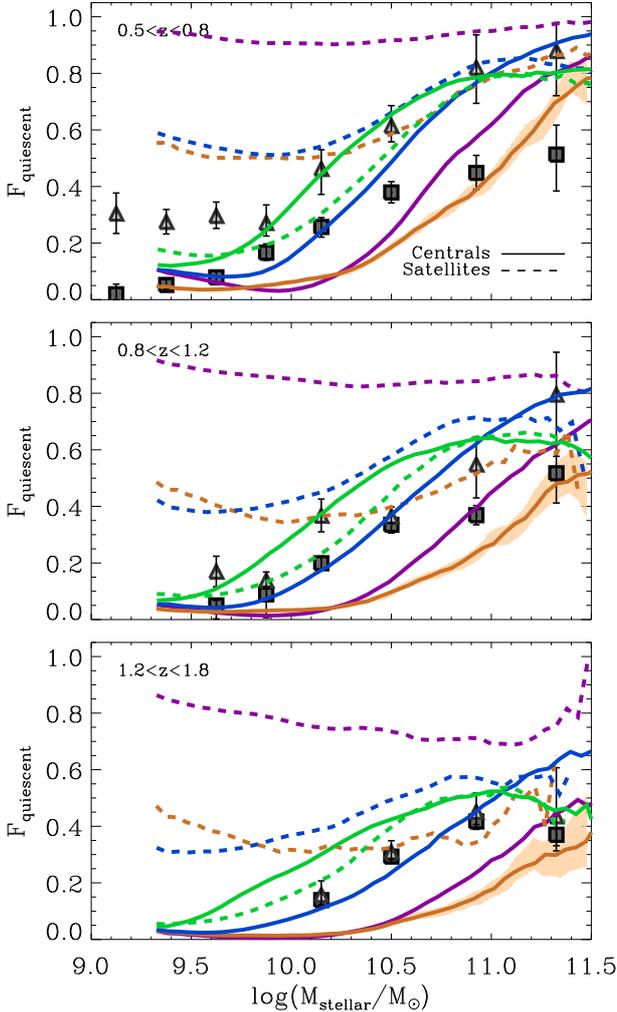,width=0.5\textwidth\hspace{-0.5cm}} 
  \caption{Fraction of passive centrals and satellite galaxies at $0.5 < z <
    1.8$. Lines of different colours correspond to the different model runs
    considered in this study (solid and dashed are used for centrals and
    satellites, respectively). Symbols with error bars (squares and triangles
    are for centrals and satellites, respectively) correspond to observational
    estimates by \citet{Fossati_etal_2017}.}  {\label{fq_highz}}
\end{figure}

Being a regime that has not been used to tune our model patameters, the high
redshift can potentially discriminate among alternative scenarios. While we
defer to a future study an accurate comparison between model predictions and
different observational measurements of the passive/quenched fraction in
different environment, we present in this section a preliminary comparison with
measurements by \citet{Fossati_etal_2017}. These authors have used
spectroscopic and grism redshifts in the five CANDELS/3D-HST fields
\citep{Grogin_etal_2011,Koekemoer_etal_2011,Brammer_etal_2012} to characterize
the environment of galaxies in the redshift range $0.5 < z <
3.0$. Multi-wavelength photometry has been used to classify galaxies as
`passive', and the fractions of passive centrals and satellites have been
evaluated as a function of stellar mass to quantify, using the same approach
used for low-z data, the fraction of galaxies whose star formation activity was
suppressed by environment specific processes. In Fig.~\ref{fq_highz}, we show
how observational estimates for central (squares) and satellite (triangles)
galaxies compare to predictions from the different runs considered above (solid
and dashed lines are used for centrals and satellites, respectively). Model
galaxies are classified as passive according to their specific SFR (see
Section~\ref{passfrac}), while a colour-colour selection has been employed for
the observational data. We have verified, however, that results are
qualitatively the same when using for model galaxies the same colour selection
adopted in Fossati et al. Results in Fig.~\ref{fq_highz} show that the fraction
of passive galaxies tend to decrease at increasing redshift, both in the models
and in the data. None of the models considered, however, reproduce well the
observed trends: the {\sc hdlf16-ed} run over-predicts significantly the
fraction of passive satellites at all redshifts considered. The passive
fractions of satellites are lower in the other runs, and the predictions from
the {\sc hdlf16-fire} model appear to be in relatively good agreement with
observational estimates for the most massive satellites ($> 10^{10}\,{\rm
  M_{\sun}}$). For central galaxies, the variation of passive fractions as a
function of galaxy stellar mass is stronger than observed in all runs
considered. At the massive end, observational data appear to be closer to
predictions from the {\sc hdlf16-fire} run, while for galaxies below $\sim
10^{10}\,{\rm M_{\sun}}$, data are close to the predictions from the {\sc
  ed-nostr1} model. For all redshift bins considered, the difference between
passive fractions for centrals and satellites remain large in all model runs
considered, while it decreases significantly in the data: at the highest
redshift the measurements for passive and centrals are very similar.

\begin{figure}
  \centering
  \epsfig{file=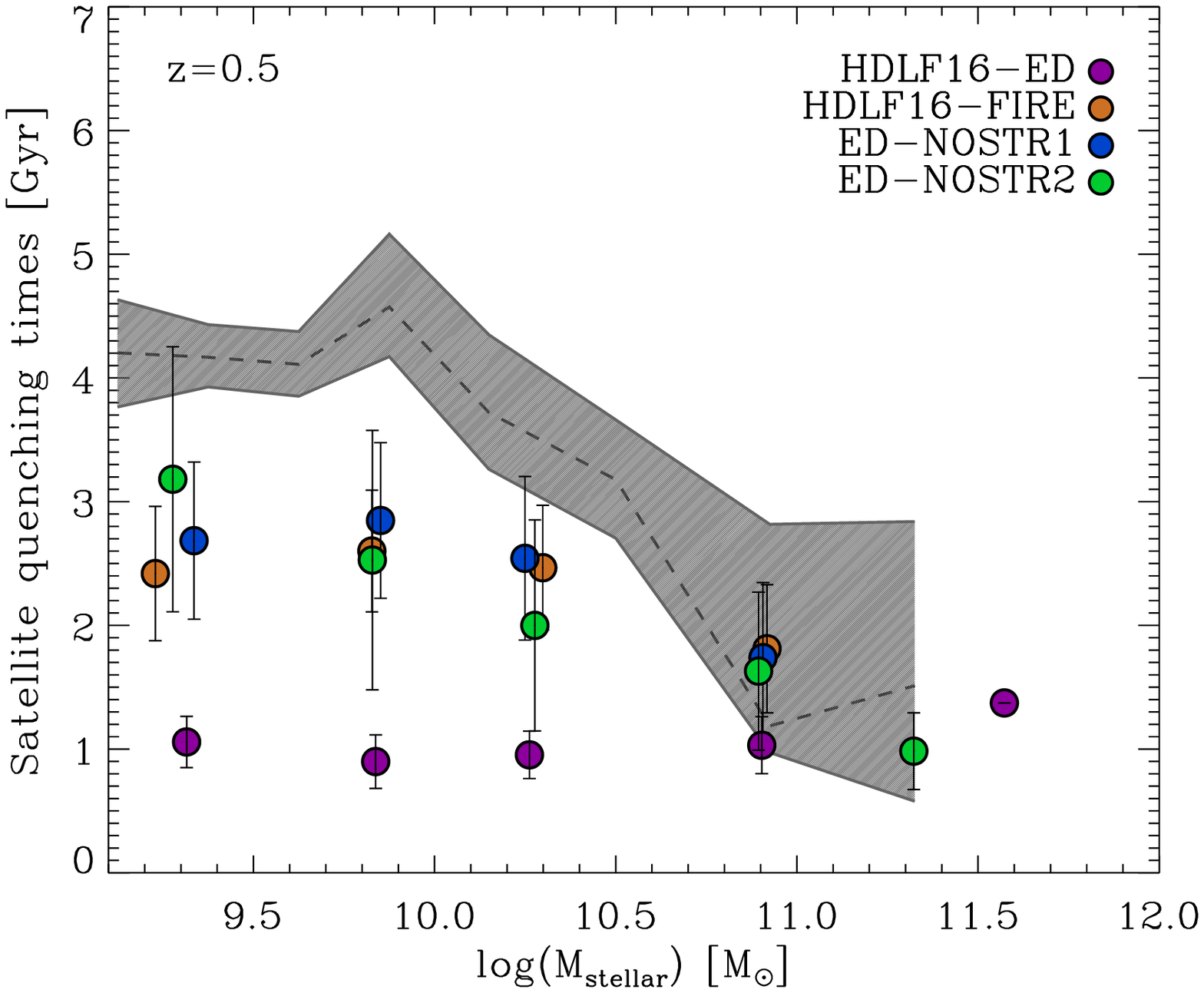,width=0.43\textwidth}
  \epsfig{file=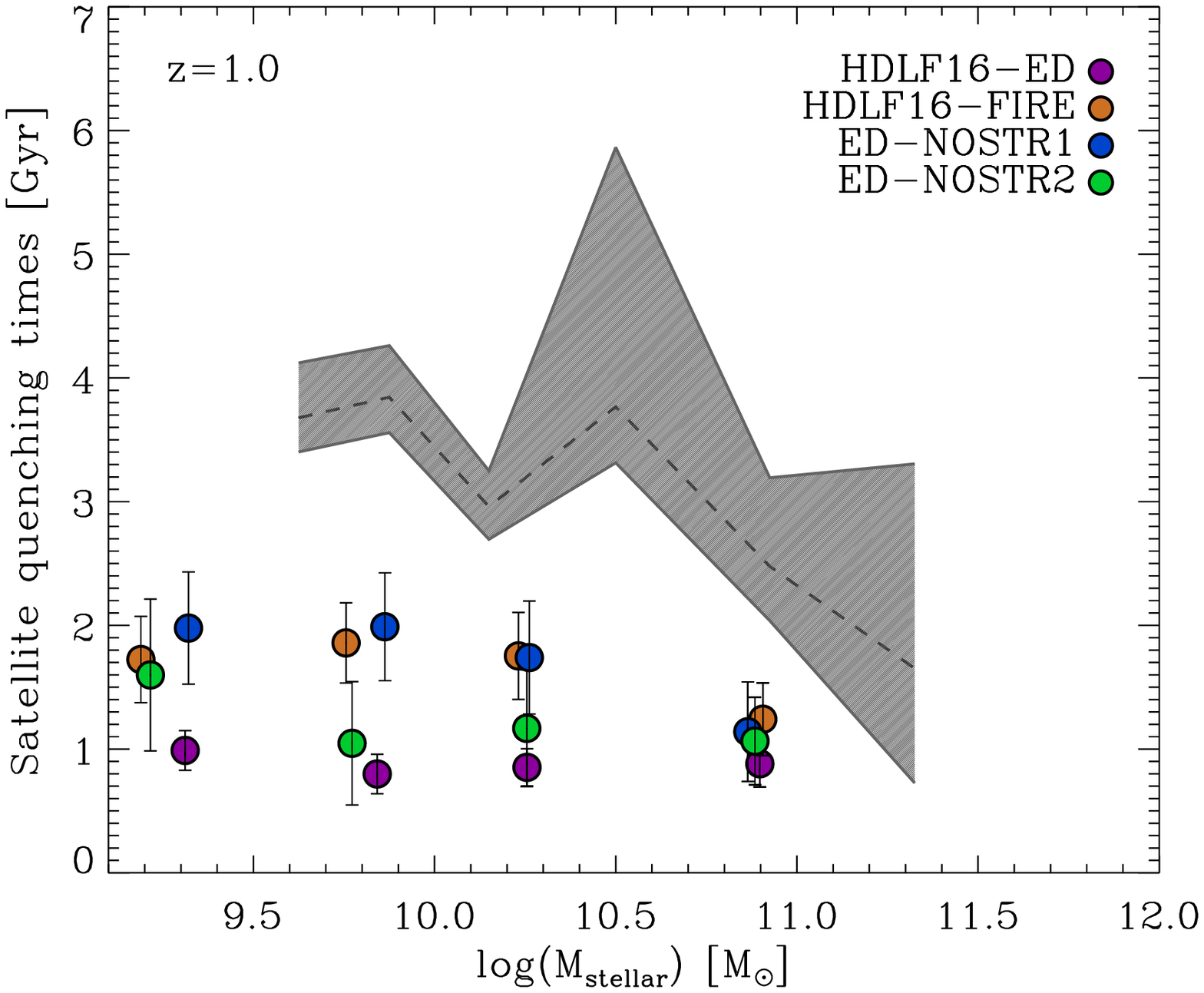,width=0.43\textwidth}
  \epsfig{file=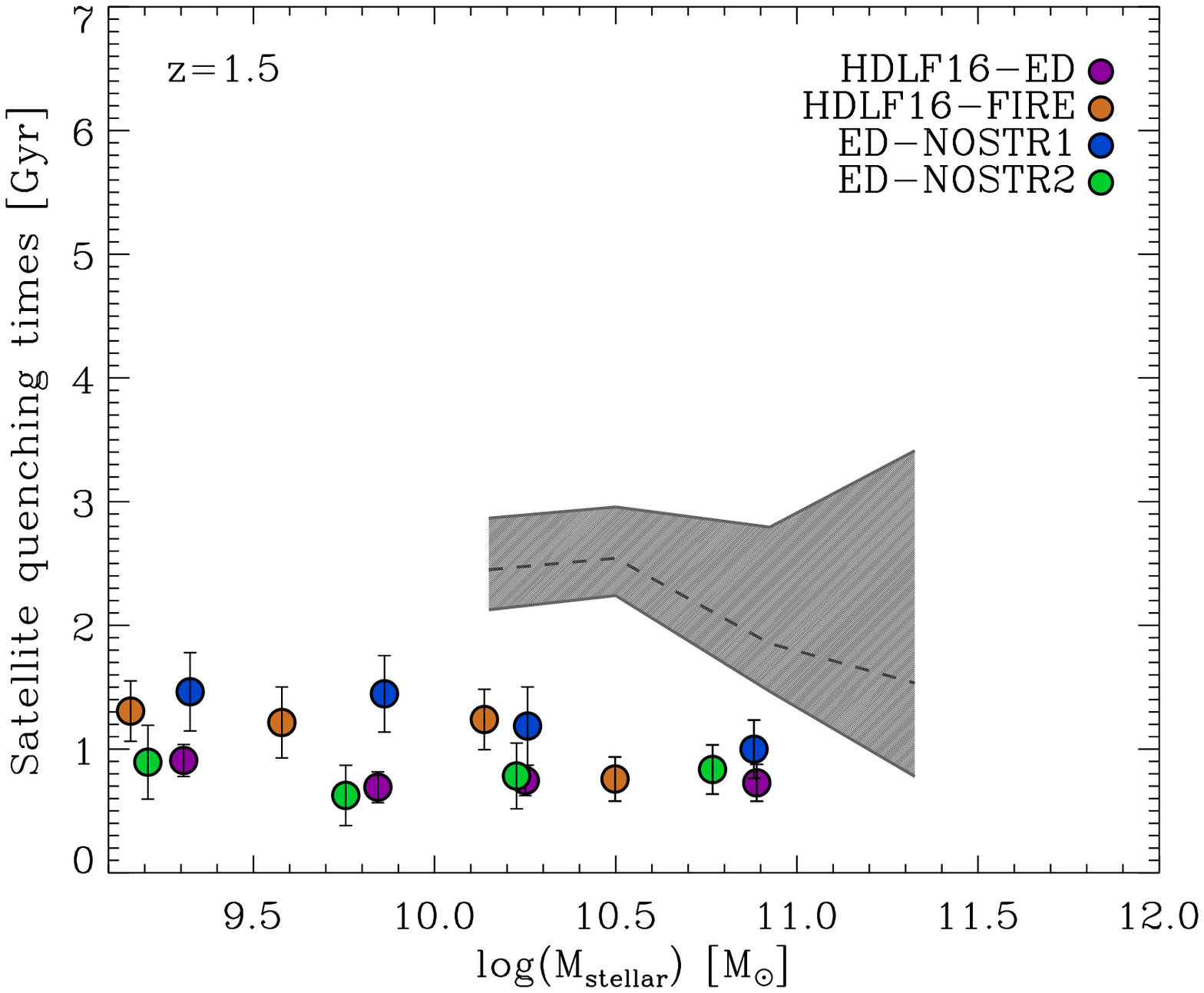,width=0.43\textwidth} 
  \caption{Quenching time-scales evaluated for passive satellites identified
    at different redshifts (different panels). Coloured symbols correspond to
    the different model runs used in this study, while dashed regions indicate
    observational estimates by \citet{Fossati_etal_2017}.}
          {\label{qt_highz}}
\end{figure}

As done in the previous section, we have also estimated the quenching
time-scales for passive satellite galaxies at different redshifts, and compared
our model predictions with observational estimates by
\citet{Fossati_etal_2017}. As said above, these are obtained using the same
approach adopted for the SDSS data. Results are shown in
Fig.~\ref{qt_highz}. Both the inferred and predicted quenching time-scales
decrease at earlier cosmic epochs, but the decrease is larger for the
theoretical predictions. At $z=0.5$, data suggest a still relatively strong
trend as a function of galaxy stellar mass. This is found also in the {\sc
  hdlf16-fire} run and in the runs assuming a non-instantaneous stripping of
the hot gas. The differences between these three runs are smaller than at
$z=0$, and the predicted quenching time-scales are lower than those inferred
from the data for galaxies with stellar mass $< 10^{10.5}\,{\rm M_{\sun}}$. At
higher redshifts, model predictions get even closer, with predicted quenching
time-scales of the order of 1~Gyr for galaxies of all masses at $z=1.5$. While
the trend as a function of galaxy stellar mass becomes less significant with
increasing redshift also in the data, the inferred quenching time-scales are
more than a factor $\sim 2$ larger than model predictions at $z=1.5$. As shown
above and in the previous section, the {\sc hdlf16-fire} model over-predicts
the fraction of passive satellites at low masses, and the disagreement becomes
more important with increasing redshift. A treatment of the non-instantaneous
stripping of the hot gas within our reference stellar feedback model can
improve the agreement with data over this mass range. We will address this in
future work, where we will also consider an explicit treatment for cold gas
removal by ram-pressure, and will carry out a more detailed comparison with
observational measurements at high redshift in different environments
\citep[e.g.][]{Balogh_etal_2016,Cucciati_etal_2017,Kawinwanichakij_etal_2017,Guo_etal_2017}.

\section{Conclusions}\label{concl} 

We have used our state-of-the-art GAlaxy Evolution and Assembly ({\sc GAEA})
semi-analytic model to analyse the characteristic time-scales of star formation
and gas consumption in satellite galaxies. In previous work, we have shown that
our fiducial model ({\sc hdlf16-fire} in this study) is able to reproduce the
measured evolution of the galaxy stellar mass function and cosmic SFR over a
significant fraction of the cosmic time
\citep{Hirschmann_etal_2016,Fontanot_etal_2017}. In addition, it exhibits a
very good agreement with important scaling relations like the mass-metallicity
relation and its evolution as a function of cosmic time
\citep{Hirschmann_etal_2016,Xie_etal_2017}. In this study, we have shown that
our fiducial model also reproduces reasonably well the variations of passive
fractions as a function of galaxy stellar mass and halo mass measured in the
local Universe, as well as the long quenching time-scales that are inferred
from data.

The same level of agreement can be obtained by using an alternative stellar
feedback scheme, that has been employed in previous versions of our model ({\sc
  hdlf16-ed}) and modifying the treatment adopted for satellite galaxies. We
find that a good agreement with the passive fractions measured in the local
Universe can be obtained only by making the unrealistic assumption that cooling
can continue on satellite galaxies, at the rate predicted using halo
  properties at the time of satellite infall, even after their parent dark
matter substructures have been stripped below the resolution limit of the
simulation, and until complete exhaustion of the hot gas ({\sc ed-nostr2}). The
stellar feedback scheme adopted in this case, however, leads to a significant
over-production of low-to-intermediate mass galaxies in our {\sc GAEA}
framework.

We demonstrate that the better agreement of the {\sc hdlf16-fire} model with
respect to previous versions of our semi-analytic model can be ascribed to (i)
a larger cold gas fraction of satellites at the time of accretion and (ii) a
reheating rate that is lower than in our older stellar feedback scheme. These
elements keep satellite galaxies active for time-scales that are significantly
longer than in previous versions of our model. A preliminary comparison with
observational estimates at higher redshift shows that our {\sc hdlf16-fire}
model reproduces well the observed passive fractions of satellites with stellar
mass larger than $\sim 10^{10}\,{\rm M}_{\sun}$, but tends to over-predict the
passive satellite fractions for lower mass galaxies and under-predict the
fraction of passive centrals.

Our results highlight the need to improve the adopted treatment for satellites
at low-mass in our {\sc hdlf16-fire} model. Given the overall better agreement
of our reference stellar feedback scheme, we argue that this is important only
for galaxy stellar masses below $\sim 10^{10}\,{\rm M}_{\sun}$. For more
massive galaxies, the abundance of passive fractions is determined primarily by
the self-regulation between star formation and stellar feedback, and is only
marginally affected by environmental processes. In future work, we plan to
include an explicit treatment for non-instantaneous stripping and for cold gas
removal by ram-pressure, and to carry out a more detailed comparison with
observational measurements at earlier cosmic epochs.

\section*{Acknowledgements}
We thank Matteo Fossati for his providing observational estimates in electronic
form, and for useful comments on a preliminary version of this paper. GDL
acknowledges financial support from the MERAC foundation.  MH acknowledges
financial support from the European Research Council via an Advanced Grant
under grant agreement no. 321323 NEOGAL.

\bibliographystyle{mnras}
\bibliography{Literaturdatenbank}
\appendix

\section{Resolution effects}
\label{app:resolution}

\begin{figure*}
  \centering
  \epsfig{file=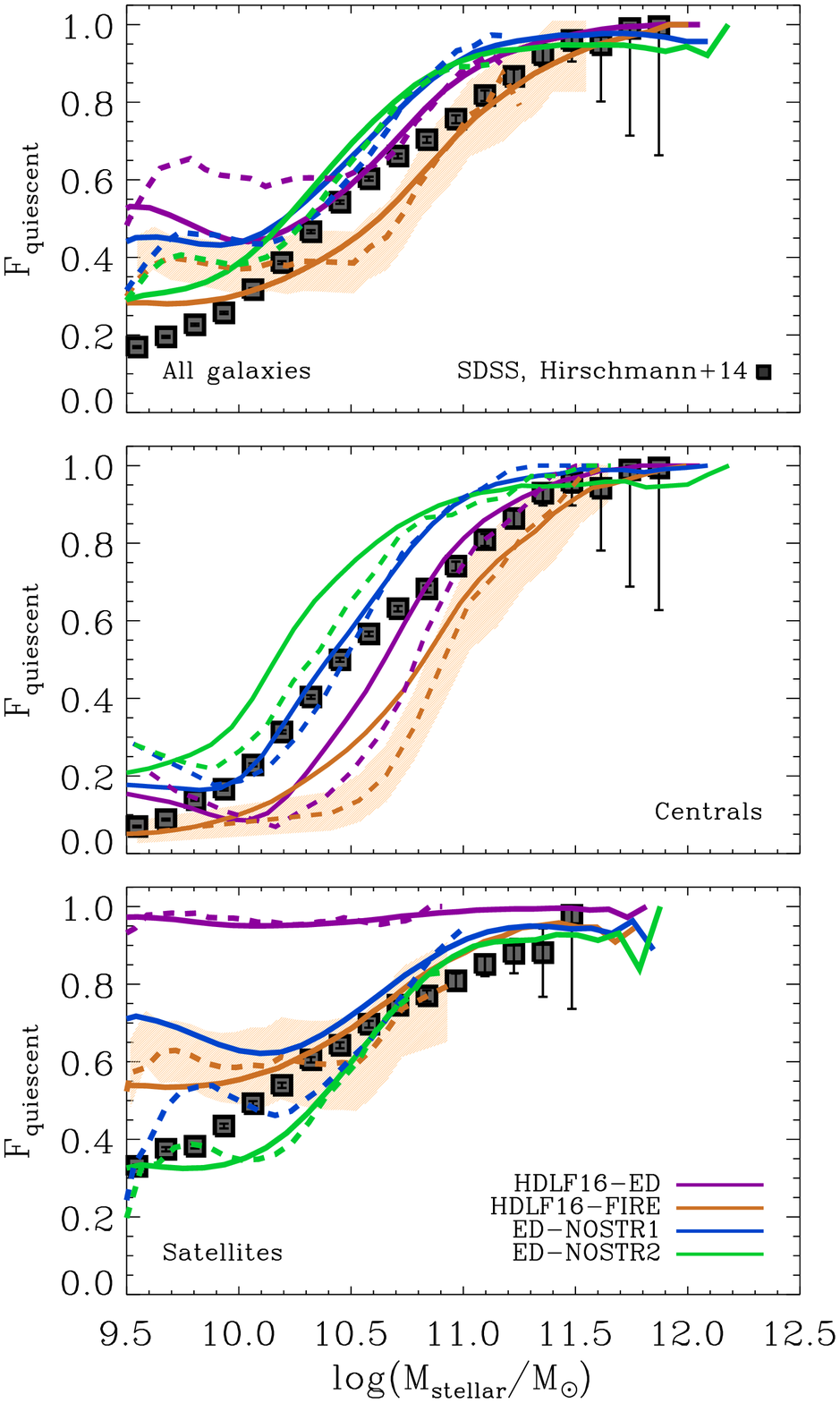,
    width=0.5\textwidth}\hspace{-0.5cm}
   \epsfig{file=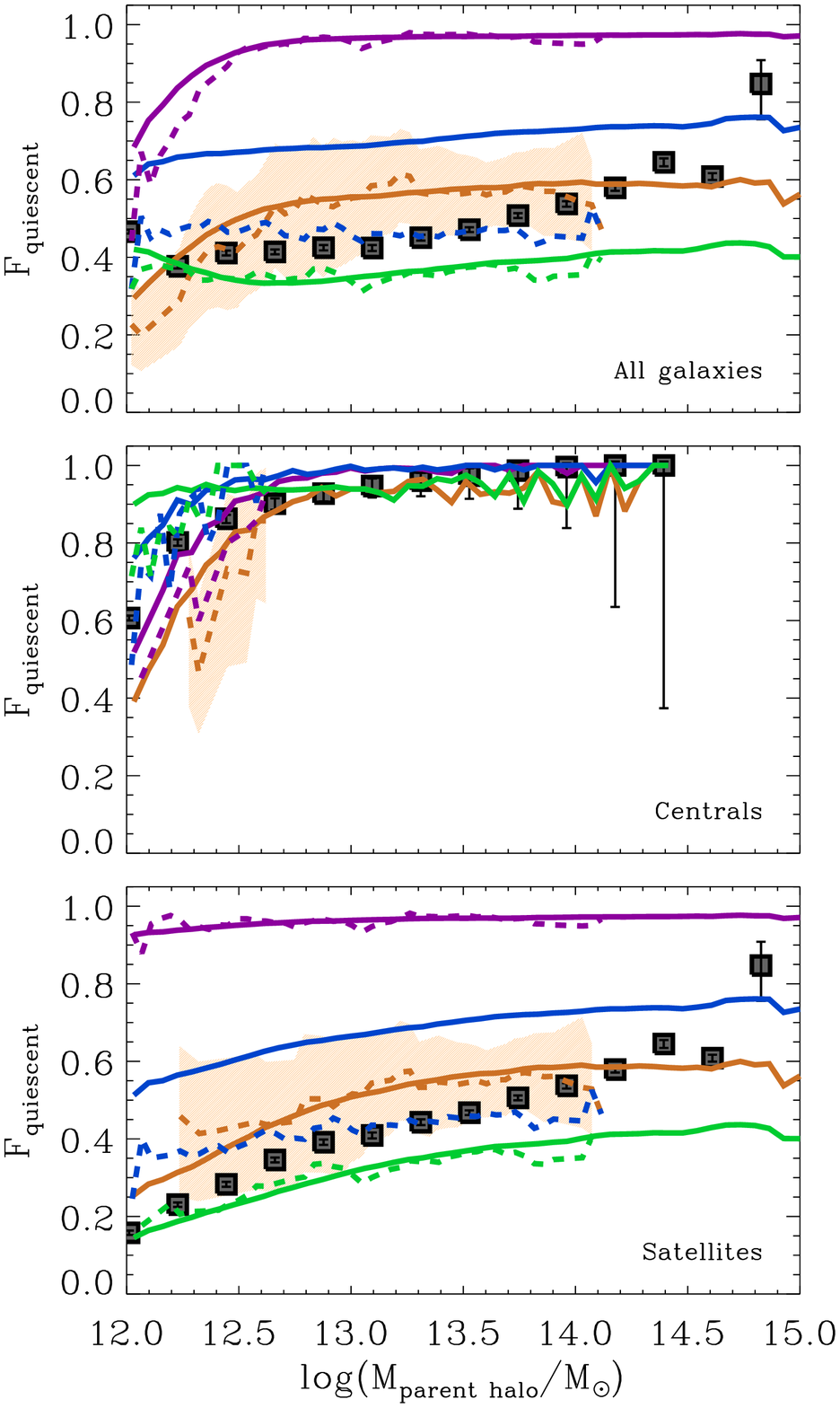,
     width=0.5\textwidth}\hspace{-0.5cm}
  \caption{As Fig.~\ref{Quiescfrac}, but including results from the
    MillenniumII Simulation as dashed lines.}  {\label{Quiescfrac_app}}
\end{figure*}

Fig.~\ref{Quiescfrac_app} shows the fraction of passive galaxies as a function
of galaxy stellar mass (left column) and parent halo mass (right column), as
predicted by all model runs considered (lines of different colours), and as
estimated from data (symbols with error bars). Solid lines correspond to model
predictions based on the Millennium Simulation (these are the same lines shown
in Fig.~\ref{Quiescfrac}), while dashed lines correspond to the Millennium II
\citep{Boylan-Kolchin_etal_2009}. Also in this case, we have used only about 20
per cent of the entire volume of the simulation, but the results shown do not
change significantly considering a larger volume. Model predictions from the
Millennium II simulation have been shown only for galaxy stellar mass bins and
halo mass bins where there are at least 30 galaxies, and only galaxies brighter
than -18 in the r-band have been considered (as for the observational data and
for model predictions based on the Millennium). For satellite galaxies (bottom
panels) model predictions are not significantly affected by resolution at the
low-mass end, except for the {\sc ed-nostr1} model. In this run, we assume that
gas can cool on satellite galaxies until there is a dark matter substructure
associated with the galaxy. At the time the subhalo is stripped below the
resolution of the simulation, we assume that the residual hot gas reservoir is
instantaneously stripped and associated with the hot reservoir of the central
galaxy. Since a higher resolution allows the substructure to be traced longer,
gas cooling can continue and keep the star formation active, moving model
predictions closer to those of the {\sc ed-nostr2} model. Although this
behaviour makes predictions from one of the models used dependent on resolution
at the low-mass end, it does not affect our main conclusions. Interestingly,
resolution also affects model predictions for central galaxies for galaxy
masses below $\sim 10^{10.5}\,{\rm M}_{\sun}$ (middle left panel), in all runs
considered in this study. Specifically, higher resolution typically translates
in lower fractions of quiescent low-mass centrals. This is likely due to the
fact that the higher resolution of the simulation leads to a larger number of
mergers with (low-mass) gas rich galaxies, which leads to bursts of star
formation. We note, however, that the difference between the two resolutions is
not very large considering the scatter (dashed orange region - it is similar
for the other models), and might be affected by the fact that we are sampling
well only relatively low-mass haloes in the Millennium II (see middle right
panel).
\label{lastpage}

\end{document}